\documentclass [12pt] {article}
\usepackage {amssymb, amsfonts}
\usepackage [T2A] {fontenc}

\usepackage [cp866] {inputenc}
\inputencoding {cp866}
\begin {document}
\vskip0.3cm

\centerline {\bf Simple estimates on size of epidemics in
generalized SIR-model} \centerline {\bf for inhomogeneous
populations} \vskip0.3cm

\centerline {{\bf E.Sh. Gutshabash {\footnote {Departament of
Physics, Sankt-Petersbourg State University, Petrodvorets,
Sankt-Petersburg, Russia; e-mail: gutshab@EG2097.spb.edu}},
M.M.Brook}} \vskip0.9cm {\footnotesize {A generalization of
Kermack-McKendick model of epidemics to the case of inhomogeneous
susceptibility of population is proposed. Some quantitative and
qualitative features of epidemic process development in this
situation are established.}}

\vskip0.9cm

The propagation of epidemics in a biological population
(collective) is a complex process. In a widely used 
Kermak-McKendric model of epidemics [1] (also known as SIR-model),
and in many others, the basic characteristic of process, the
susceptibility coefficient $\lambda$, is supposed to be constant
over a population, i.e. all individuals are supposed to have the
same immunity. Numerous researches, however, show that this
assumption is not satisfied. For various reasons wide fluctuations
of susceptibility parameters are observed in real populations, and
they may influence significantly the dynamics of epidemic process.
In the present work we generalize the SIR-model to the case of
inhomogeneous susceptibility (immunity) of the population.

To obtain the equations of the model we shall assume that a closed
population of $N$ individuals can be split into $n$ groups ($2 \le
n \le N $), such that all individuals from the $i$-th group have
the same susceptibility coefficient $ \lambda_i> 0, \:i=1, n $.
The system of equations of Kermack-McKendrick for this case has
the form (the dot stands for the time derivative):

$$ \dot S_i =-\lambda_iS_iI, \:\:\:\:\dot I=I\sum _ {i=1}
^n\lambda_iS_i-\gamma I, \:\:\: \:\dot R =\gamma I, \eqno (1) $$
where $S_i=S_i (t) $ is the number of individuals in the $i$-th
group susceptible to an infection at the moment $t $, $I=I (t) $
is the number of infected at the moment $t $, $R=R (t) $ is the
number of recovered at the moment $t $, and $ \gamma> 0 $ is the "
factor of elimination". The initial conditions for the system (1)
have the form:

$$
S_i (0) =S _ {0i}, \:\:\:\:I (0) =I_0, \:\:\:\:R (0 =0. \eqno (2)
$$

Notice that in model (1), similar to elementary models of the
chemical kinetics, we neglect the spatial distribution of
individuals.

To estimate the effect of inhomogeneity of the immunity we are
going to compare the characteristics of the inhomogeneous
population under consideration and of the corresponding
homogeneous population with a susceptibility given by

$$ \lambda _ {hom} = {\bar \lambda} = \frac {\sum _ {i=1} ^n
\lambda_iS _ {0i}} {S_0}, \eqno (3) $$ where $S_0 =\sum _ {i=1}
^nS _ {0i} $ is the total number of susceptible individuals at
$t=0 $. The average susceptibility is constant for homogeneous
populations, whereas in the inhomogeneous case it is easily seen
to be a decreasing function of time:

$$ {\dot {\bar \lambda}} _ {nonhom} = I [(\frac {\sum _ {i=1} ^n
\lambda_iS_i} {\sum _ {i=1} ^nS_i}) ^2- \frac {\sum _ {i=1}
^n\lambda_i^2S_i} {\sum _ {i=1} ^nS_i}] <0, \eqno (4) $$ which
significantly influences the process. In the initial moment we
have

$$ \dot S _ {nonhom} (0) - \dot S (0) =-I_0\sum _ {i=1}
^n\lambda_iS _ {0i} + {\bar \lambda} I_0S_0=0, $$ $$ \eqno (5) $$
$$ \ddot S _ {nonhom} (0)-\ddot S_0=I_0 [\sum _ {i=1}
^n\lambda_i^2S _ {i0} - (\sum _ {i=1} ^n\lambda_iS _ {0i}) ^2] <0,
$$ where $ \dot S (0), \:\ddot S (0), \:\dot S _ {nonhom} (0),
\:\ddot S _ {nonhom} (0) $ are the "speed" \enskip and
"acceleration" of the process at $t=0 $ for the homogeneous and
inhomogeneous populations, respectively.

Thus, at $t=0 $ the speeds are equal, but the accelerations are
different, and $ | \ddot S _ {nonhom} (0) | <| \ddot S (0) | $.
Hence, the inequality (4) implies that at the initial stage the
epidemic process in the inhomogeneous population develops slower
than in the homogeneous one.

From (1) we have the following "conservation law":

$$
\sum _ {i=1} ^nS _ {0i} e ^ {-\lambda_i\int_0^tI (\tau) d\tau} +I (t) + \gamma\int_0^tI (\tau) d\tau=N.
\eqno (6)
$$

Let $s ^ {*} =\int_0 ^ {\infty} I (\tau) d\tau $. This quantity is
related to the size of epidemics: $z (\infty) = \gamma s ^ {*} $,
and, under the assumption that $I (\infty) =0 $, is a root of the
equation

$$ N-\gamma s ^ {*}-\sum _ {i=1} ^n S _ {0i} e ^ {-\lambda_is ^
{*}} =0. \eqno (7) $$ The quantities $s ^ {*} $ and $z (\infty) $
are continuous functions of parameters $ \lambda_i $, and $s ^ {*}
$ varies from $I_0/\gamma $ to $N/\gamma $ monotonically in $
\lambda_i $.

We are now going to find approximative (asymptotic) solutions to
the transcendent equation (7). To this end, we shall consider
three ranges of values of parameters.

1. In the range

$$
\lambda_is ^ {*} \gg 1       \eqno(8)
$$
we have:

$$ s ^ {*} \cong \frac {N} {\gamma} \:\:\: {\mbox and} \:\:\: z
(\infty) \cong N, \eqno (9) $$ that is, the epidemic will finally
get the whole population.

2. In the range

$$ \lambda_is ^ {*} \cong 1       \eqno(10) $$ we get

$$ s ^ {*} \cong \frac {1} {\gamma} (N-\frac {S_0} {e}) \eqno (11)
$$ and, correspondingly,

$$
\lambda_1 =\lambda_1 =\ldots = \lambda_n \cong \frac {\gamma} {N-\frac {S_0} {e}},  \eqno(12)
$$
$$
z (\infty) \cong N-\frac {S_0} {e}. \eqno (13)
$$

3. In the range

$$ \lambda_is ^ {*}\ll 1,       \eqno(14) $$ developing the
exponentials in (7) up to the second order terms, we find:

$$ z (\infty) \cong \frac {2I_0\gamma} {\gamma-\sum _ {i=1} ^nS _
{0i} \lambda_i +\sqrt {(\gamma-\sum _ {i=1} ^n S _ {0i} \lambda_i)
^2+2I_0\sum _ {i=1} ^n S _ {0i} \lambda_i^2}}. \eqno (15) $$
Hence, in this range, the inhomogeneity of the population can
affect the size of epidemics. In turn, the size of epidemics in
the corresponding homogeneous population with the susceptibility $
{\bar \lambda} $ has the form:

$$ z _ {hom} (\infty) \cong \frac {2I_0\gamma} {\gamma-S_0 {\bar
\lambda} + \sqrt {(\gamma-S_0 {\bar \lambda}) ^2+2I_0S_0 {\bar
\lambda} ^2}}. \eqno (16) $$ It is not difficult to show using the
Silvester criterion that

$$ z (\infty) <z _ {\hom} (\infty), \eqno (17) $$ that is, the
size of epidemics in the homogeneous population is greater than in
the inhomogeneous one.

Let us suppose now that the population is large and the quantity $
\lambda $ is a random variable. We shall assume that in the limit
$n \to N $ the actual susceptibility of the population can be
described by some non-negative continuous probability density
$f(\lambda)$,

$$ \int_0 ^ {\infty} f (\lambda) d\:\lambda=1 . \eqno (18) $$
Let

$$
E\lambda =\lambda ^ {*} =\int_0 ^ {\infty} \lambda f (\lambda) d\lambda \eqno (19)
$$
and

$$ \sigma^2=E\lambda^2-(E\lambda) ^2 \eqno (20) $$ be,
respectively, the expectation and the dispersion of
susceptibility. The estimate (15) then becomes:

$$
z (\infty) \cong \frac {2I_0\gamma} {b (\lambda ^ {*}) +\sqrt {(b (\lambda ^ {*})) ^2+2I_0a (\lambda ^ {*},\sigma^2)}}, \eqno (21)
$$
where

$$ \:\:\:a (\lambda ^ {*}) = S_0 (\lambda ^ {*2} + \sigma^2),
\:\:\:\:b (\lambda ^ {*}) =\gamma-S_0\lambda ^ {*}. \eqno (22) $$
Notice that the integration in (14) - (15) is taken over the
positive real axis, for the case $ \lambda <0 $ corresponds to
individuals absolutely resistant to infection, which, as a rule,
are non-existent in real populations.

Let us find now the influence of the dispersion of susceptibility
on the size of epidemics. Let $ \lambda ^ {*} $ be fixed. It then
follows from expressions (21)-(22), that the size of epidemics
grows as $ \sigma^2 $ decreases and reaches the maximum $$ \lim _
{\sigma^2 \to 0} z (\infty) = \frac {2I_0\gamma}
{\gamma-S_0\lambda ^ {*} +\sqrt {(\gamma-S_0 \lambda ^ {*}) ^
2+2I_0S_0\lambda ^ {*2}}}, \eqno (23) $$ which is consistent with
the discrete case (11).

Let's pass to the managing parameter $ \rho =\gamma/\lambda ^ {*}$ and we shall analyse the expression (23) at $I_0 \ll (\rho-S_0) ^2 $. The relation for the size of epidemie then becomes more simple:

$$
z (\infty) \cong \frac {-\rho [\rho-S_0-| S_0-\rho |]} {S_0}. \eqno (24)
$$
In the considered area of parameter's values two cases are possible:

1. $ \rho <S_0 $, that is carried out
at enough small $I_0 $, i.e. at small number of ills. In this situation

$$
z (\infty) \cong 2\rho (1-\frac {\rho} {S_0}) \eqno (25)
$$
and
$$
\lim _ {S_0 \to \infty} z (\infty) =2\rho. \eqno (26)
$$
Thus, the size of epidemic in case of the big population with small initial number of ills
will be limited and to be determined by the quantity of the managing parameter.

2. $S_0 <\rho $. Then for the size of epidemic we shall obtain

$$
z (\infty) =0. \eqno (27)
$$

So, within the framework of the considered model it was possible to show, that heterogeneity of a population on a susceptibility (immunity) influences on epidemic process.

Restrictions on the received by us in the given work areas it is possible to interpret, if
to take into account the expression for average relative speed epidemic "waves":

$$
v_i =\frac {-\int_0 ^ {\infty} \frac {\dot S_i} {S_i} dt} {T_i}, \eqno (28)
$$
where $T_i $ is the duration of epidemie in $i$-th group. Then the expression (10), determining
the second area, it is possible to replace on the expression

$$
{\bar v} = \frac {1} {T}, \eqno (29)
$$
where $T$ is the duration of epidemie in a population, ${\bar v}$  is average speed of the epidemie "waves".

For relative speed, using the first equation of system (1), we shall obtain:

$$
{\bar v} +\lambda_0I_0=0, \eqno (30)
$$
where
$$
\lambda_0 =\frac {\gamma} {N-\frac {S_0} {e}}, \:\:\:v =\frac {\dot S} {S}. \eqno (31)
$$
Then for the first area

$$
{\bar v} \gg \frac {1} {T}, \:\:\: {\bar v} + \lambda_0I_0 <0, \eqno (32)
$$
and, similarly, for the third area

$$
{\bar v} <\frac {1} {T}, \:\:\: {\bar v} + \lambda_0I_0> 0. \eqno (33)
$$
The relations (30)-(33) are convenient that allow to easily estimate the size of epidemie even at the initial moment of time on the value of size $ {\bar v} + \lambda_0I_0 $.

In summary, we shall note, that in view of complexity and an insufficient level of scrutiny of the mechanism of immunity, the corresponding mathematical models, and also, based on them, model of epidemies are, certainly, enough rough. Nevertheless, apparently, even the estimations received in the elementary models, in practice can be appear rather effective.

\vskip1cm

1. W.O.Kermack and A.G. McKendrick. Proc. R.Soc. Edinburg A 115, 700 (1927).

\end{document}